\documentclass[prl,longbibliography,twocolumn,superscriptaddress,amsmath,amssymb,verbatim]{revtex4-1}

\usepackage{graphicx}
\usepackage{lmodern}
\usepackage{epstopdf}
\usepackage[colorlinks=true,citecolor=blue,linkcolor=blue,urlcolor=blue,bookmarks=true]{hyperref}
\usepackage{xcolor}

\newcommand{\YCu}{YCu$_3$(OH)$_6$Cl$_3$}

\begin{document}

\preprint{APS/123-QED}

\title{Origin of Magnetic Ordering in a Structurally Perfect Quantum Kagome Antiferromagnet}

\author{T.~Arh}
\affiliation{Jo\v{z}ef Stefan Institute, Jamova c.~39, SI-1000 Ljubljana, Slovenia}
\author{M.~Gomil\v{s}ek}
\affiliation{Jo\v{z}ef Stefan Institute, Jamova c.~39, SI-1000 Ljubljana, Slovenia}
\affiliation{Centre for Materials Physics, Durham University, South Road, Durham, DH1 3LE, UK}
\author{P.~Prelov\v{s}ek}
\affiliation{Jo\v{z}ef Stefan Institute, Jamova c.~39, SI-1000 Ljubljana, Slovenia}
\author{M.~Pregelj}
\affiliation{Jo\v{z}ef Stefan Institute, Jamova c.~39, SI-1000 Ljubljana, Slovenia}
\author{M.~Klanj\v{s}ek}
\affiliation{Jo\v{z}ef Stefan Institute, Jamova c.~39, SI-1000 Ljubljana, Slovenia}
\author{A.~Ozarowski}
\affiliation{National High Magnetic Field Laboratory, Florida State University, Tallahassee, Florida 32310, USA}
\author{S.~J.~Clark}
\affiliation{Centre for Materials Physics, Durham University, South Road, Durham, DH1 3LE, UK}
\author{T.~Lancaster}
\affiliation{Centre for Materials Physics, Durham University, South Road, Durham, DH1 3LE, UK}
\author{W.~Sun}
\affiliation{Fujian Provincial Key Laboratory of Advanced Materials, Department of Materials Science and Engineering, College of Materials, Xiamen University, Xiamen 361005, Fujian Province, People's Republic of China}
\author{J.-X.~Mi}
\affiliation{Fujian Provincial Key Laboratory of Advanced Materials, Department of Materials Science and Engineering, College of Materials, Xiamen University, Xiamen 361005, Fujian Province, People's Republic of China}
\author{A.~Zorko}
\email{andrej.zorko@ijs.si}
\affiliation{Jo\v{z}ef Stefan Institute, Jamova c.~39, SI-1000 Ljubljana, Slovenia}
\affiliation{Faculty of Mathematics and Physics, University of Ljubljana, Jadranska u.~19, SI-1000 Ljubljana, Slovenia}

\date{\today}

\begin{abstract}
The ground state of the simple Heisenberg nearest-neighbor quantum kagome antiferromagnetic model is a magnetically disordered spin liquid, yet various perturbations may lead to fundamentally different states.
Here we disclose the origin of magnetic ordering in the structurally perfect kagome material YCu$_3$(OH)$_6$Cl$_3$, which is free of the widespread impurity problem.
{\it Ab initio} calculations and modeling of its magnetic susceptibility  reveal that, similar to the archetypal case of herbertsmithite, the nearest-neighbor exchange is by far the dominant isotropic interaction.
Dzyaloshinskii-Moriya (DM) anisotropy deduced from electron spin resonance, susceptibility and specific-heat data is, however, significantly larger than in herbertsmithite. 
By enhancing spin correlations within kagome planes, this anisotropy is essential for magnetic ordering.
Our study isolates the effect of DM anisotropy from other perturbations and unambiguously confirms the predicted phase diagram. 
\end{abstract}

\maketitle

\begin{figure}[b]
\includegraphics[trim = 0mm 0mm 0mm 0mm, clip, width=1\linewidth]{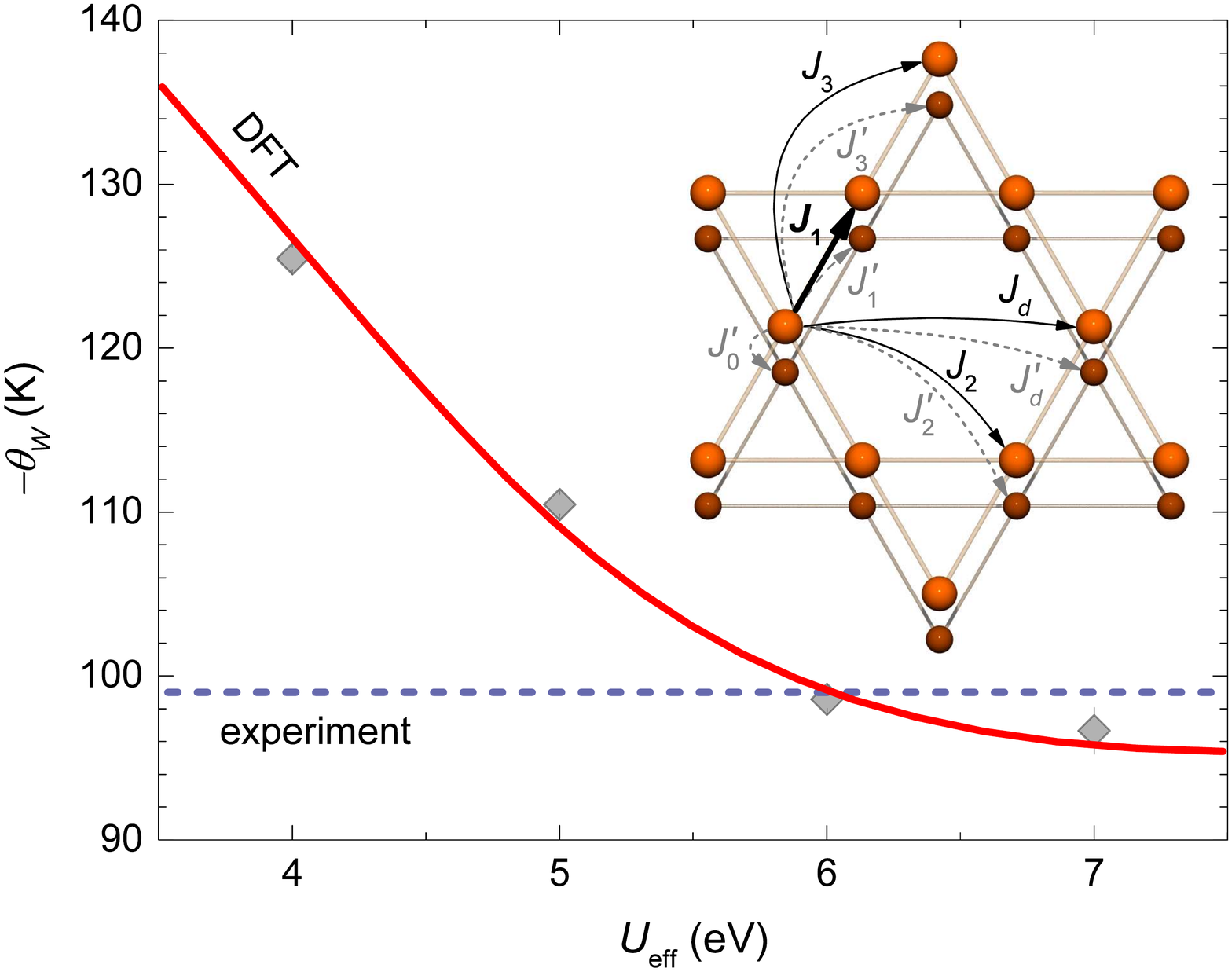}
\caption{Weiss temperature $\theta_{W}$ of {\YCu} determined from DFT+$U$ calculations for different values of the effective on-site Hubbard repulsion $U_{\rm eff}$ (points).
The dashed line shows the experimental value $\theta_{W}=-99$\,K, while the solid line serves as a guide to the eye.
The inset depicts two neighboring kagome layers of Cu$^{2+}$ spin-1/2 ions with inplane Heisenberg exchange interactions $J_i$ (solid arrows) and interplane interactions $J'_i$ (dashed arrows).
The nearest-neighbor coupling $J_1$ is by far the dominant one \cite{sup}.}
\label{fig1}
\end{figure}

Quantum spin liquids are magnetically disordered, yet highly entangled states, promoted by quantum fluctuations on some geometrically frustrated spin lattices \cite{lacroix2011introduction}.
A paradigm predicting such a state even at zero temperature is the two-dimensional (2D) nearest-neighbor quantum kagome antiferromagnetic model (KAFM) \cite{balents2010spin,savary2017quantum,zhou2017quantum}, represented by Heisenberg, i.e., isotropic $J_1$ exchange bonds between spins-1/2 sites in Fig.\,\ref{fig1}.
Yet, even small perturbations to this simple model can stabilize fundamentally different ground states, as their influence is strongly amplified by frustration. 
Various factors, including further-neighbor exchange interactions \cite{suttner2014renormalization, gong2015global,  bieri2016projective,gotze2016route,hering2017functional,zhu2019identifying, bernu2019effect}, magnetic anisotropy \cite{hering2017functional,zhu2019identifying,bernu2019effect,cepas2008quantum,chernyshev2014quantum,
rousochatzakis2009dzyaloshinskii}, defects \cite{rousochatzakis2009dzyaloshinskii,singh2010valence,kawamura2014quantum}, and structural distortions \cite{norman2019valence} have been the focus of theoretical investigations in recent years.
One of the seminal predictions that still calls for a clear experimental validation is a quantum critical point induced by Dzyaloshinskii-Moriya (DM) magnetic anisotropy, separating a spin liquid from a magnetically ordered ground state of KAFM \cite{cepas2008quantum}.
Here we elucidate the role of the DM interaction in promoting correlations that lead to magnetic ordering in a material that closely realizes the KAFM.

Actual KAFM realizations are as a rule plagued by several perturbations, making the assessment of the individual roles of these perturbations challenging.
A direct consequence of many effects being intertwined is that even the existence of a spin gap in the spin-liquid ground state of the KAFM remains unsettled.
In fact, for the hitherto most intensively studied KAFM material herbertsmithite \cite{norman2016herbertsmithite}, indications of a finite gap \cite{fu2015evidence} have been recently superseded by the conclusion that the gap is absent \cite{khuntia2020gapless}.
However, the effects of particular perturbations present in this material on its low-energy magnetism remain unknown.
Relevant imperfections include sizable inter-site ion mixing \cite{de2008magnetic,olariu200817,freedman2010site}, large DM anisotropy \cite{zorko2008dzyaloshinsky} and subtle structural distortion away from perfect kagome symmetry \cite{zorko2017symmetry,laurita2019evidence}. 
On the contrary, in the recently synthesized KAFM material {\YCu}  \cite{sun2016perfect} 
no structure-related perturbations are present; there is no Cu--Y inter-site disorder \cite{sun2016perfect} and the initially reported small Y-site disorder \cite{sun2016perfect} is absent in high-resolution neutron diffraction of high-quality powder samples \cite{berthelemy2019local}.
Therefore, the recent discoveries of static internal magnetic fields below $T_N=12$\,K \cite{zorko2019YCu3muon,berthelemy2019local} and magnetic Bragg peaks at low temperatures \cite{zorko2019negative} are rather surprising.
Initially, a broad maximum in specific heat at a notably higher temperature of $T_{\rm max} = 16$\,K was also assigned to 3D ordering \cite{zorko2019YCu3muon}, causing a discrepancy with $T_N$ where static internal fields appear.
Experiments have further established that the average ordered Cu$^{2+}$ magnetic moment of an otherwise regular $120^{\circ}$ magnetic structure is strongly reduced 
\cite{zorko2019negative} and is accompanied by persisting
spin fluctuations even at the lowest temperatures \cite{zorko2019YCu3muon}.
The origin of such exotic magnetism is unknown, but even more fundamentally, the basic question of the magnetic-ordering mechanism present in this material remains unexplained.
Since {\YCu} is a unique KAFM material with a very limited number of possible perturbations, determining the ordering mechanism would be very important for  assessing the impact of these perturbations on the spin-liquid ground state of KAFM.

Here we show a combination of density functional theory (DFT), finite-temperature Lanczos method (FTLM) and electron spin resonance (ESR) results,
which allows us to address the origin of the unexpected magnetic ordering in {\YCu}.
DFT calculations together with modeling of the magnetic susceptibility show that the nearest-neighbor Heisenberg exchange $J_1=82(2)$\,K is by far the dominant isotropic interaction. 
Almost perfect agreement between numerical modeling and complementary ESR measurements, magnetic susceptibility and specific heat data reveals an additional sizable out-of-plane DM anisotropy $D_z/J_1=0.25(1)$ that places the investigated compound in the magnetically ordered region of the KAFM phase diagram \cite{cepas2008quantum}.
Moreover, FTLM modeling provides a novel insightful view into the role of DM interaction in KAFM and allows the precise determination of $D_z$, 
which is responsible for the maximum in specific heat at $T_{\rm max}=16$\,K related to the enhancement of 2D chiral spin correlations.
3D order is established via a small inter-layer exchange  below $T_N=12$\,K, where static internal magnetic fields appear \cite{zorko2019YCu3muon}.  

\begin{figure}[t]
\includegraphics[trim = 0mm 0mm 0mm 0mm, clip, width=1\linewidth]{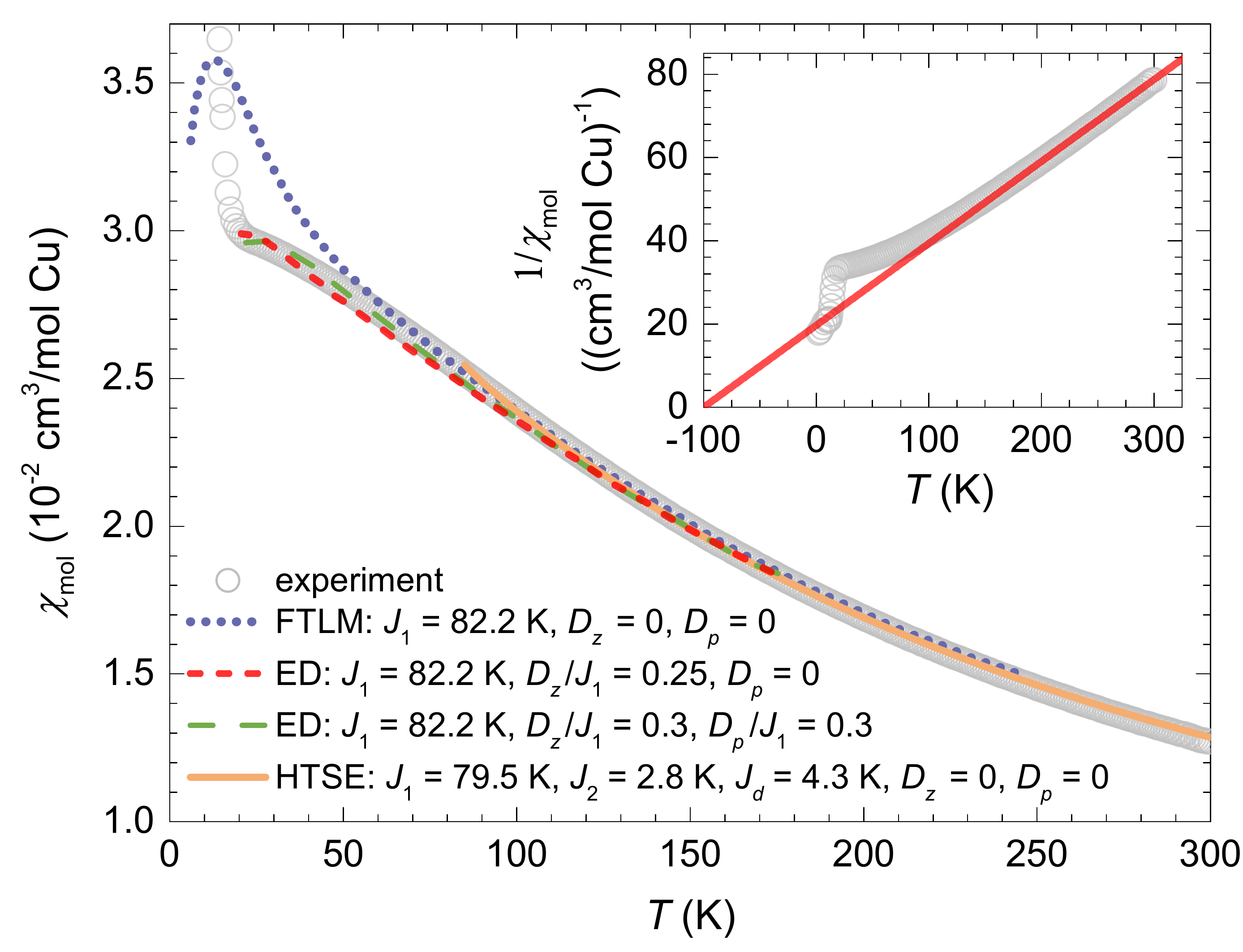}
\caption{Molar susceptibility $\chi_{\rm mol}$ of {\YCu} in a field of 0.1\,T \cite{zorko2019YCu3muon}, with its sharp increase at low temperatures indicating magnetic ordering. 
The solid line is a fit with the HTSE $J_1$--$J_2$--$J_d$ model \cite{bernu2013exchange}. 
The dotted line shows FTLM calculations for isotropic KAFM on $N=42$ sites 
\cite{schnack2018magnetism}. 
The dashed lines are ED calculations with additional out-of-plane DM component $D_z$ and in-plane component $D_p$ for $N=15$ sites \cite{rigol2007kagome}, which are accurate to within 4\% down to $T_N$ \cite{sup}.
The inset shows a Curie-Weiss analysis, $1/\chi_{\rm mol}=(T-\theta_{W})/C$, with the Weiss temperature $\theta_{W}=-99$\,K and $g$ factor $g=2.077$.
}
\label{fig2}
\end{figure}

To understand the magnetism of {\YCu}, the first task is to determine its dominant isotropic exchange interactions.
As in other kagome compounds \cite{janson2008modified, jeschke2013first, iqbal2015paramagnetism, jeschke2015barlowite}, we tackle this problem using total-energy (broken-symmetry) DFT+$U$ calculations \cite{riedl2019ab} (for details see Ref.\,\cite{sup}).
We assume that each site is coupled with sites up to the third nearest neighbor in the kagome layer and with equivalent sites in the neighboring two kagome layers (Fig.\,\ref{fig1}).
Our calculated exchange constants and the corresponding Weiss temperature  
$\theta_W =-\sum_i z_i J_i/4$,
where $z_i$ is the number of neighbors coupled to a particular site with  $J_i$ \cite{goodenough1963magnetism}, depend on the effective on-site Hubbard repulsion $U_{\rm eff}$ \cite{sup}.
$\theta_W$ is compared with its experimental value of -99(1)\,K, which is obtained from a Curie-Weiss fit to the susceptibility data (inset in Fig.\,\ref{fig2}).
The experiment is well reproduced for $U_{\rm eff}= 6$\,eV (Fig.\,\ref{fig1}), a value consistent with previous studies on similar materials \cite{janson2008modified, jeschke2013first, jeschke2015barlowite, iqbal2015paramagnetism}.
We find that the exchange interaction between nearest neighbors $J_1=84.2(4)$\,K by far exceeds all other Heisenberg interactions, as all of them are below 5\% of $J_1$, irrespective of the chosen value of $U_{\rm eff}$ \cite{sup}.

Next, we focus on the temperature dependence of the magnetic susceptibility to verify that the calculated exchange constants are consistent with experiment.
We first compare the experimental susceptibility \cite{zorko2019YCu3muon} to a high temperature series expansion (HTSE) calculation for a simplified $J_1$--$J_2$--$J_d$ model \cite{bernu2013exchange} in Fig.\,\ref{fig2}.
The HTSE curve fitted in the temperature range between 100 and 300\,K matches the experiment very well and yields the exchange constants $J_1=79.5(1)$\,K, $J_2=2.8(27)$\,K, and $J_d=4.3(54)$\,K.
Furthermore, we can compare the experiment to FTLM calculations for a pure nearest-neighbor KAFM on a $N=42$ spin cluster \cite{schnack2018magnetism}.
Good agreement is obtained for temperatures down to $0.6J_1\sim 50$\,K with $J_1=82.2(1)$\,K being the only free parameter (Fig.\,\ref{fig2}).

The fact that all three independent approaches yield very similar predictions, namely a dominant Heisenberg exchange interaction $J_1=82(2)$\,K, gives strong credibility to these results.
As isotropic exchange interactions beyond the nearest neighbors are limited to at most 5\% of $J_1$, {\YCu} can be placed alongside herbertsmithite \cite{jeschke2013first} as one of the best realizations of the nearest-neighbor KAFM.
In all other well-studied examples, like kapellasite \cite{janson2008modified,bernu2013exchange, iqbal2015paramagnetism}, haydeeite \cite{janson2008modified,boldrin2015haydeeite, iqbal2015paramagnetism}, volborthite \cite{janson2016magnetic}, and vesignieite \cite{boldrin2018vesignieite}, further-neighbor interactions are much larger.
As interactions $|J_2|,\,|J_3|,\,|J_d|\gtrsim 0.2J_1$ \cite{suttner2014renormalization, gong2015global,  bieri2016projective, hering2017functional} or $|J'|\gtrsim 0.15J_1$ \cite{gotze2016route} are needed to induce magnetic ordering in the KAFM, these are evidently too small in {\YCu}.
The only remaining perturbation that can account for its ordered ground state is magnetic anisotropy.
Since there are no symmetry restrictions \cite{moriya1960anisotropic},
both the antisymmetric DM and the symmetric anisotropic exchange (AE) interaction are allowed.
However, DM anisotropy is generally dominant in Cu$^{2+}$-based magnets because it is a one order lower correction to the isotropic exchange  \cite{moriya1960anisotropic}, so that the AE term is smaller by a factor $\Delta g/g \sim 0.2$ \cite{abragam1970electron}. 
Here $\Delta g$ is the shift of the $g$ factor from the free electron value. 
Furthermore, as the easy-plane AE interaction that would be compatible with the observed planar magnetic order \cite{zorko2019negative} does not lead to ordering of the KAFM \cite{chernyshev2014quantum, he2015distinct}, 
we expect the DM interaction to play a dominant role.

The next task is, therefore, to determine the DM interaction ${\bf D} \cdot ({\bf S}_i\times {\bf S}_j)$ between the nearest neighbors.
First, we note that further-neighbor isotropic exchange interactions are too small to account for the large discrepancy between the experimental magnetic susceptibility and the nearest-neighbor FTLM calculations
already at temperatures as high as $0.6J_1\sim 50$\,K (Fig.\,\ref{fig2}).
On the contrary, a sizable DM interaction can explain this deviation.
Indeed, according to exact-diagonalization (ED) calculations \cite{rigol2007kagome}, the out-of-plane component $D_z$ suppresses susceptibility compared to the isotropic KAFM, while the in-plane component $D_p$ enhances it \cite{sup}.
The experimental suppression 
is well reproduced for $D_z/J_1=0.25(1)$ all the way down to the ordering temperature if $D_p = 0$ (Fig.\,\ref{fig2}).
For $D_p > 0$ a larger $D_z$ is required \cite{sup}, e.g., for $D_p/J_1 = 0.30$  one finds $D_z/J_1 = 0.30(1)$ (Fig.\,\ref{fig2}).

\begin{figure}[t]
\includegraphics[trim = 0mm 0mm 0mm 0mm, clip, width=1\linewidth]{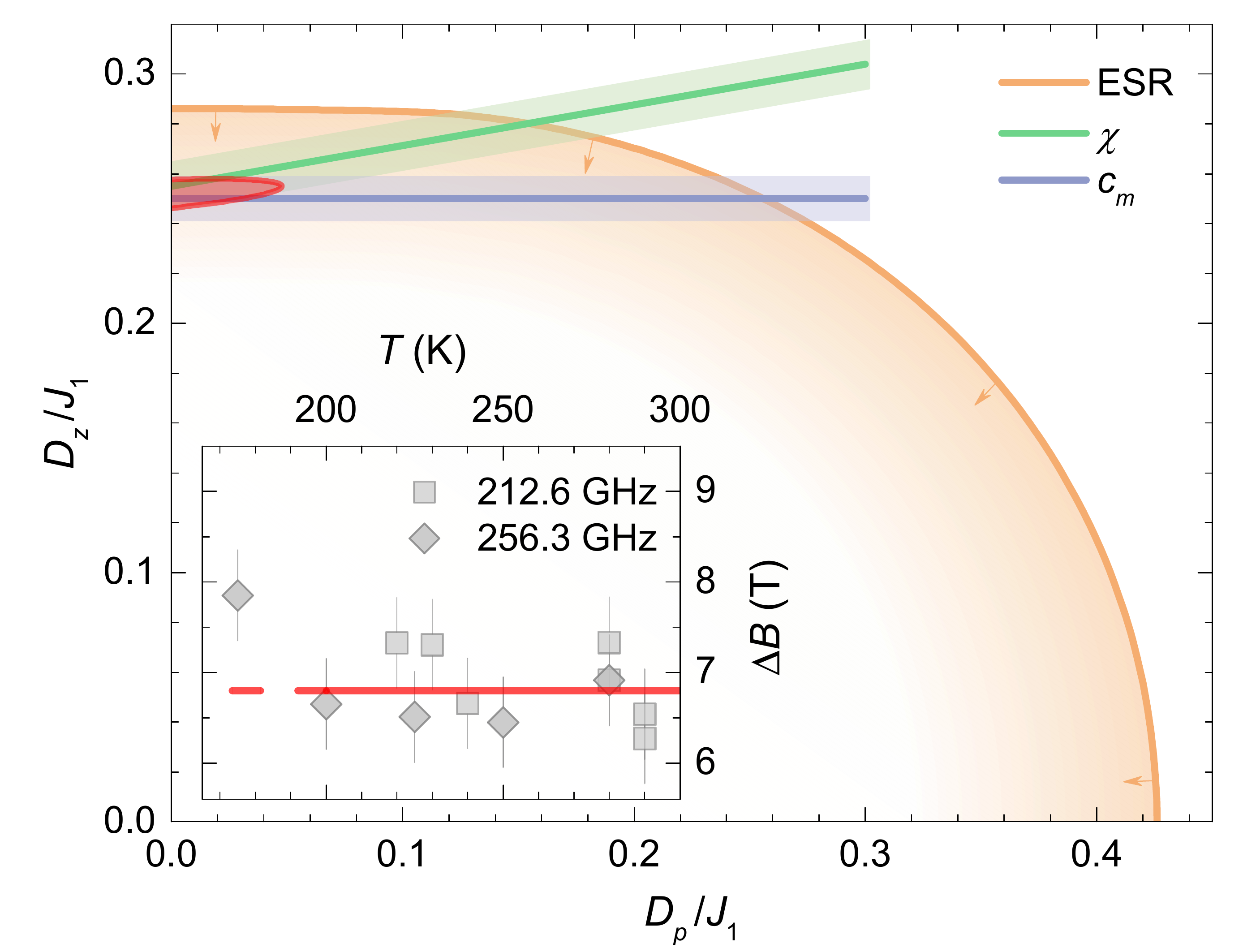}
\caption{The interdependence of both DM components in {\YCu} based on the analysis of the ESR linewidth (shown in the inset), magnetic susceptibility and specific heat.
Shaded regions show experimental uncertainty, while the arrows imply that ESR only gives an upper bound. 
The red area is the region with globally acceptable parameters, where the solid red line indicates the 1-sigma boundary.
}
\label{fig3}
\end{figure}

We can place further constraints on the magnitude of both DM components based on ESR results
(for details see Ref.\,\cite{sup}), as magnetic anisotropy directly broadens the ESR spectra \cite{zorko2018determination}.
The measured spectra \cite{sup} are broader than in other Cu-based kagome compounds like herbertsmithite \cite{zorko2008dzyaloshinsky}, vesignieite \cite{zorko2013dzyaloshinsky}, and kapellasite \cite{kermarrec2014spin} by almost an order of magnitude.
Above 200\,K the ESR linewidth is constant at $\Delta B = 6.8(5)$\,T (inset in Fig.\,\ref{fig3}), which is consistent with the high-temperature paramagnetic regime and allows for the application of Kubo-Tomita (KT) theory \cite{kubo1954general}.
The well-established expression for the ESR linewidth on the kagome lattice \cite{zorko2008dzyaloshinsky,zorko2013dzyaloshinsky} allows us to derive the $D_z (D_p)$ solution \cite{sup} shown in Fig.\,\ref{fig3}.
Contrary to the case of susceptibility, which is affected oppositely by the two DM components, they both broaden the ESR linewidth. 
The total magnitude of the DM vector is therefore approximately limited by
$D/J_1 \simeq \left[2g \mu_B \Delta B/( \sqrt{\pi}k_B J_1) \right]^{1/2}=0.36,$
where $k_B$ is the Boltzmann constant and $\mu_B$ is the Bohr magneton.
The joint ESR and susceptibility analysis yields the limits $0.25<D_z/J_1<0.29$ and $D_p/J_1<0.15$ (Fig.\,\ref{fig3}).   
We note, though, that in accordance with recent ED calculations demonstrating  that the KT approach might somewhat overestimate the DM anisotropy on the kagome lattice \cite{el2010electron}, the true DM components should be closer to the lower limits, $D_z/J_1=0.25$ and $D_p/J_1\simeq 0$.

\begin{figure*}[t]
\includegraphics[trim = 0mm 0mm 0mm 0mm, clip, width=1\linewidth]{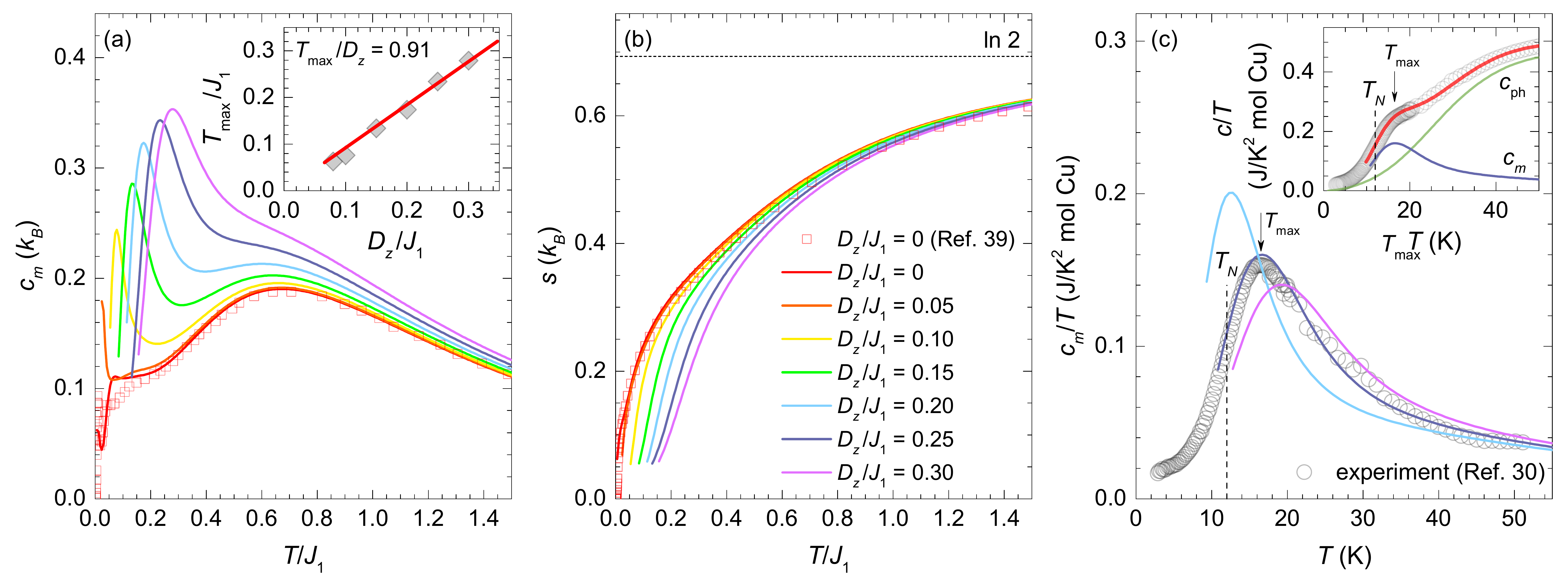}
\caption{
The temperature dependence of (a) the magnetic specific heat $c_m$ and (b) the entropy per site $s$ in zero magnetic field obtained from the FTLM calculations  on $N=30$ spin clusters for various $D_z$ values (lines). 
The data are shown only for temperatures where $s\geq 0.07 k_B$ \cite{sup}.
The results from Ref.\,\cite{schnack2018magnetism} obtained on $N=42$ spin clusters for $D_z/J_1=0$ are shown for comparison (symbols).
The inset in (a) shows the variation of the low-temperature $c_m$ maximum position $T_{\rm max}$ with $D_z/J_1$.
(c) Selected FTLM calculations of $c_m/T$ compared to the experimental data (circles).
The inset shows the total specific heat of {\YCu} (circles; data taken from Ref.\,\cite{zorko2019YCu3muon}) 
and the fit (red line) composed of the magnetic contribution for $D_z/J_1=0.25$ (blue line) and a phonon contribution (green line) \cite{sup}.
The arrow indicates $T_{\rm max}=16$\,K and the dashed line $T_N=12$\,K.
}
\label{fig4}
\end{figure*}

An independent check of the above estimates is provided by modeling  previously published zero-field specific heat ($c$) data \cite{zorko2019YCu3muon}.
FTLM calculations \cite{jaklic94finite,jaklic00finite} of the magnetic contribution to the specific heat $c_m$, which were performed on spin clusters with up to $N=30$ spins for various $D_z/J_1$ and $D_p/J_1$ ratios (for details see Ref.\,\cite{sup}),
reveal two well-resolved maxima in $c_m$ for $D_z/J_1 \gtrsim 0.08$ (Fig.\,\ref{fig4}a), 
as previously also observed in ED calculations on smaller clusters \cite{rigol2007kagome}.
A broad high-temperature maximum is, similarly to the spin-1/2 square lattice \cite{sengupta2003specific}, found around $0.67J_1$ and does not shift with the DM interaction. 
Therefore, it is associated with the enhancement of nearest-neighbor spin correlations \cite{chakravarty1988low, chakravarty1989two}. 
On the contrary, a much
 narrower low-temperature maximum shifts almost linearly with the out-of-plane DM component and is found at $T_{\rm max}\simeq 0.91 D_z$ (inset in Fig.\,\ref{fig4}a).
In sharp contrast, $c_m$ is almost insensitive to the in-plane DM component at least up to $D_p/J_1 \leq 0.3$ \cite{sup, rigol2007kagome}.
As the $D_z$ term linearly shifts the energy of the 120$^\circ$ spin structure of basic kagome triangles \cite{zorko2019negative} while $D_p$ does not, we attribute the low-temperature maximum to growing chiral spin correlations within the kagome planes. 
This makes specific heat a unique probe of the DM component $D_z$ on the kagome lattice, which is much more sensitive (Fig.\,\ref{fig4}c) than magnetic susceptibility \cite{sup}.

For $D_z/J_1=0.25$, the predicted magnetic specific heat nicely matches the experiment (Fig.\,\ref{fig4}c).
Indeed, we can fit the $c/T$ data very well with the model $c=c_m+c_{\rm ph}$ that includes a phonon contribution $c_{\rm ph}$. 
The fit is already good for a simple Debye phonon model with the
Debye temperature $\theta_D = 224(5)$\,K and is further improved by including an additional Einstein phonon contribution (inset in Fig.\,\ref{fig4}c) \cite{sup}, corresponding to a Raman active mode at 123\,cm$^{-1}$, as found in structurally similar herbertsmithite \cite{wulferding2010interplay}.   
The obtained $D_z/J_1=0.25(1)$ is in excellent agreement with the lower-bound estimate based on ESR and susceptibility modeling (Fig.\,\ref{fig3}) and thus provides further evidence that the in-plane DM component is much smaller, i.e., $D_p/J_1<0.05$.
Although the DM anisotropy in {\YCu} is larger than in some other Cu$^{2+}$-based KAFM materials \cite{zorko2008dzyaloshinsky, zorko2013dzyaloshinsky}, its size is  compatible with the order-of-magnitude estimate \cite{moriya1960anisotropic} $D_z/ J_1\sim \Delta g/g \sim 0.2$ for the Cu$^{2+}$ ions \cite{abragam1970electron}.

Having established the main terms in the spin Hamiltonian of {\YCu}, we are now in position to discuss the origin of its magnetic ordering.
It is theoretically well established that the out-of-plane DM interaction leads to a $q= 0$ long-range order of KAFM at zero temperature if its strength exceeds the critical value $D_z^c = 0.10(2) J_1$ \cite{cepas2008quantum,rousochatzakis2009dzyaloshinskii,
hering2017functional,zhu2019identifying} separating the spin liquid and the ordered phase.
Contrary to the paradigmatic KAFM material herbertsmithite, which appears to be on the verge of criticality \cite{zorko2008dzyaloshinsky}, we find that {\YCu} lies well inside the ordered phase.
Nevertheless, the average ordered moment should be strongly suppressed due to quantum fluctuations.
Indeed, the predicted moment of 0.35\,$\mu_B$ for $D_z/J_1=0.25$ \cite{cepas2008quantum} matches reasonably well with the experimental value of 0.42(2)\,$\mu_B$ \cite{zorko2019negative}. 

Finally, let us comment on the compatibility of our results with the celebrated Mermin--Wagner theorem \cite{mermin1966absence}, which precludes long-range order in the considered 2D model at nonzero temperatures
due to continuous in-plane symmetry.
As revealed by FTLM calculations, 2D short-range chiral order is established below $T_{\rm max}=16$\,K, while 3D order is only established below $T_N= 12$\,K (Fig.\,\ref{fig4}c), where static internal magnetic fields appear, longitudinal muon spin relaxation rate suddenly starts increasing and bulk susceptibility exhibits a clear cusp (see Fig.\,6 in Ref.\,\cite{sup}).
Finite $T_N$ requires additional interlayer interactions $J'$ and is determined by the growth of the in-plane correlation length $\xi$ to the extent that the thermal energy drops below the interaction energy of short-range ordered 2D regions on neighboring kagome planes, when $T_N\approx [\xi(T_N)/d]^2 J'S(S+1)$, with $d$ being the nearest-neighbor distance \cite{chakravarty1988low, chakravarty1989two}.
As $\xi$ should only marginally depend on the interlayer interaction for $J'/J_1\ll1$ and thus $T_N$ should only logarithmically depend on $J'$  \cite{chakravarty1989two,yasuda2005neel,sengupta2003specific}, $T_N$ is dominantly determined by $D_z$ in {\YCu}.
This anisotropy promotes building up of 2D chiral spin correlations, which corresponds to effectively shifting a large release of the system's entropy to temperatures around $T_{\rm max}\approx D_z$ (Fig.\,\ref{fig4}b).
As a result, for $J'/J_1\ll 1$ yielding $T_N<T_{\rm max}$ most of the entropy is already released around $T_{\rm max}$ and the effective number of degrees of freedom involved in 3D ordering is significantly reduced, making the cusp in $c_m$ at $T_N$ unobservable \cite{sengupta2003specific}.
Thus, 2D physics essentially prevails down to $T_N$ and justifies the absence of any cluster-size dependence of the $c_m$ curves in FTLM calculations \cite{sup}.
    

In conclusion, {\YCu} turns out to be an extremely rare structurally perfect KAFM material, with by far the nearest-neighbor isotropic exchange interaction  $J_1=82(2)$\,K dominating all other isotropic interactions, while by far the most relevant perturbation is the out-of-plane DM anisotropy $D_z/J_1 = 0.25(1)$.  
This is determined from a perfect coincidence of the experiments and numerical calculations for the two most common bulk magnetic characterization techniques as well as ESR, which is unique in the field of frustrated magnetism.
Such $D_z/J$ places the system in the magnetically ordered part of the predicted phase diagram \cite{cepas2008quantum}. 
This provides an unambiguous experimental confirmation of the key role of the DM interaction in inducing magnetic order on the kagome lattice.
Furthermore, now that this role is well understood, a  sister compound Y$_3$Cu$_9$(OH)$_{18}$OCl$_8$ with a slightly distorted kagome lattice and apparently a spin-liquid ground state \cite{berthelemy2019local} provides an ideal opportunity to study the effects of further perturbations.  
Since in this compound very similar exchange interactions and magnetic anisotropy as in {\YCu} are expected, the reasoning for its lack of magnetic ordering should be searched in deviations from perfect kagome symmetry.
  
The authors thank O.~C\'epas, F.~Bert and P.~Mendels for fruitful discussions.
This work was supported by the Slovenian Research Agency under Projects No.~BI-US/18-20-064, No.~Z1-1852, and Program No.~P1-0125.
M.G., T.L. and S.J.C. are grateful to EPSRC (UK) for financial support through Grant No.~EP/N024028/1.
The National High Magnetic Field Laboratory is supported by National Science Foundation through NSF/DMR-1644779 and the State of Florida.
Computing resources were provided by the F1 Department at IJS, STFC Scientific Computing Department's SCARF cluster and the Durham HPC Hamilton cluster. Research data from the UK effort will be made available via Durham Collections.

%

%

%
\newpage
\begin{widetext}
\vspace{19cm}
\begin{center}
{\large {\bf Supplemental Information: Origin of Magnetic Ordering in a Structurally Perfect Quantum Kagome Antiferromagnet}}\\
\vspace{0.5cm}
T.~Arh,$^{1}$ M.~Gomil\v sek,$^{1, 2}$ P.~Prelov\v sek,$^{1}$ M.~Pregelj,$^{1}$ M.~Klanj\v sek,$^{1}$ A.~Ozarowski,$^{3}$\\
S.~J.~Clark,$^{2}$ T.~Lancaster,$^{2}$ W.~Sun,$^{4}$ J.-X.~Mi,$^{4}$ and A.~Zorko$^{1, 5, *}$
\vspace{0.3cm}

{\it \small
$^1$Jo\v{z}ef Stefan Institute, Jamova c.~39, SI-1000 Ljubljana, Slovenia\\
\vspace{0cm}
$^2$Centre for Materials Physics, Durham University, South Road, Durham, DH1 3LE, UK\\
\vspace{0cm}
$^3$National High Magnetic Field Laboratory, Florida State University, Tallahassee, Florida 32310, USA\\
\vspace{0cm}
$^4$Fujian Provincial Key Laboratory of Advanced Materials, Department of Materials Science and Engineering, College of Materials, Xiamen University, Xiamen 361005, Fujian Province, People's Republic of China\\
\vspace{0cm}
$^5$Faculty of Mathematics and Physics, University of Ljubljana, Jadranska u.~19, SI-1000 Ljubljana, Slovenia\\
}

\end{center}
\end{widetext}
\section{DFT Calculations}

\textit{Ab-initio} calculations were performed using the CASTEP DFT code~\cite{clark2005firstSI} using a local (spin) density approximation (LDA) functional with an additional effective on-site Hubbard repulsion $U_{\rm eff} = U - J_H = 4$--$7~\mathrm{eV}$. 
Here, $U$ is the bare Hubbard repulsion and $J_H\approx 1~\mathrm{eV}$ is Hund's coupling, which was kept fixed. An LDA+$U$ functional was chosen over a more sophisticated generalized gradient approximation (GGA) functional, as recent results suggest~\cite{sharma2018sourceSI} that it can, in certain cases, perform better than standard GGA+$U$ functionals~\cite{perdew1996generalizedSI} when describing the magnetism of materials.

For each value of $U_{\rm eff}$ the starting $173(2)~\mathrm{K}$ experimental crystal structure from Ref.\,\cite{sun2016perfectSI} was relaxed with free lattice in internal structural parameters, resulting in unit cells with $93(1)\%$ of the experimental cell volume. 
Well-converged total DFT energies of $102$ random collinear spin configurations in a $2 \times 2 \times 2$ supercell containing $24$ $\mathrm{Cu}^{2+}$ ions were then calculated and finally fitted in the total-energy (broken-symmetry) DFT framework~\cite{riedl2019abSI} by a spin model described by $9$ Heisenberg exchange interactions depicted in Fig.\,1 of the main text plus an overall energy offset. 
The results of these fits as well as the derived Weiss temperatures $\theta_W =-\sum_i z_i J_i/4$ are summarized in Table\,\ref{tab:exchange}.
We note that even though the true spin ground state of {\YCu} is noncollinear \cite{zorko2019negativeSI}, the dominant exchange couplings extracted from collinear total-energy DFT calculations are still expected to be reliable, as was also found in other frustrated systems \cite{riedl2019abSI,janson2008modifiedSI, jeschke2013firstSI, iqbal2015paramagnetismSI, jeschke2015barlowiteSI}. 
The reason for this is that we are concerned with parametrizing the isotropic part of the system's spin Hamiltonian, for which knowing the energies of
(excited-state) collinear spin configurations is sufficient.

%
\begin{table}[t]
 \centering
 \caption{Isotropic exchange coupling constants of {\YCu} calculated by DFT+$U$ for various values of the effective on-site Hubbard repulsion $U_{\rm eff}$.
 The last column corresponds to HTSE modeling of magnetic susceptibility (see Fig.\,2 in the main text). 
The considered inplane constants $J_i$ and interplane constants $J'_i$ are defined in Fig.\,1 in the main text.
 The Weiss temperature is $\theta_W =-\sum_i z_i J_i/4$,
where $z_i$ denotes the number of neighbors coupled to a particular site by $J_i$ \cite{goodenough1963magnetismSI}.}
  \begin{tabular*}{\linewidth}{@{\extracolsep{\fill}} c  c c c c | c}
   \hline \hline
$U_{\rm eff}$\,(eV) & 4  & 5 & 6 & 7 & HTSE\\ 
\hline
$J_1$\,(K) & 107.3(5) & 94.2(4) & 84.2(4) & 85.2(4) & 79.5(1) \\
$J_2$\,(K) & 5.2(5) & 4.1(5) & 3.7(4) & 3.0(4) & 2.8(27) \\
$J_3$\,(K) & 4.2(4) & 3.2(3) & 2.6(3) & 1.8(3) & / \\
$J_d$\,(K) & 4.5(6) & 3.8(5) & 3.6(4) & 2.6(4) & 4.3(54)\\
$J'_0$\,(K) & 0.9(5) & 0.7(4) & 0.7(4) & 0.1(4) & / \\
$J'_1$\,(K) & -0.2(3) & 0.1(2) & -0.2(2) & 0.0(2) & / \\
$J'_2$\,(K) & 2.0(3) & 1.9(3) & 1.9(2) & 1.7(2) & / \\
$J'_3$\,(K) & 0.3(1) & 0.4(1) & 0.3(1) & 0.3(1) & / \\
$J'_d$\,(K) & 2.0(2) & 2.0(2) & 1.9(2) & 1.6(2) & /\\
\hline
$-\theta_W$\,(K) & 125.5(7) & 110.5(6) & 98.6(5) & 96.7(14) & / \\
   \hline \hline
 \end{tabular*}
 \label{tab:exchange}
\end{table}

\section{Modeling of magnetic susceptibility}

\begin{figure*}[t]
\includegraphics[trim = 0mm 0mm 0mm 0mm, clip, width=1\linewidth]{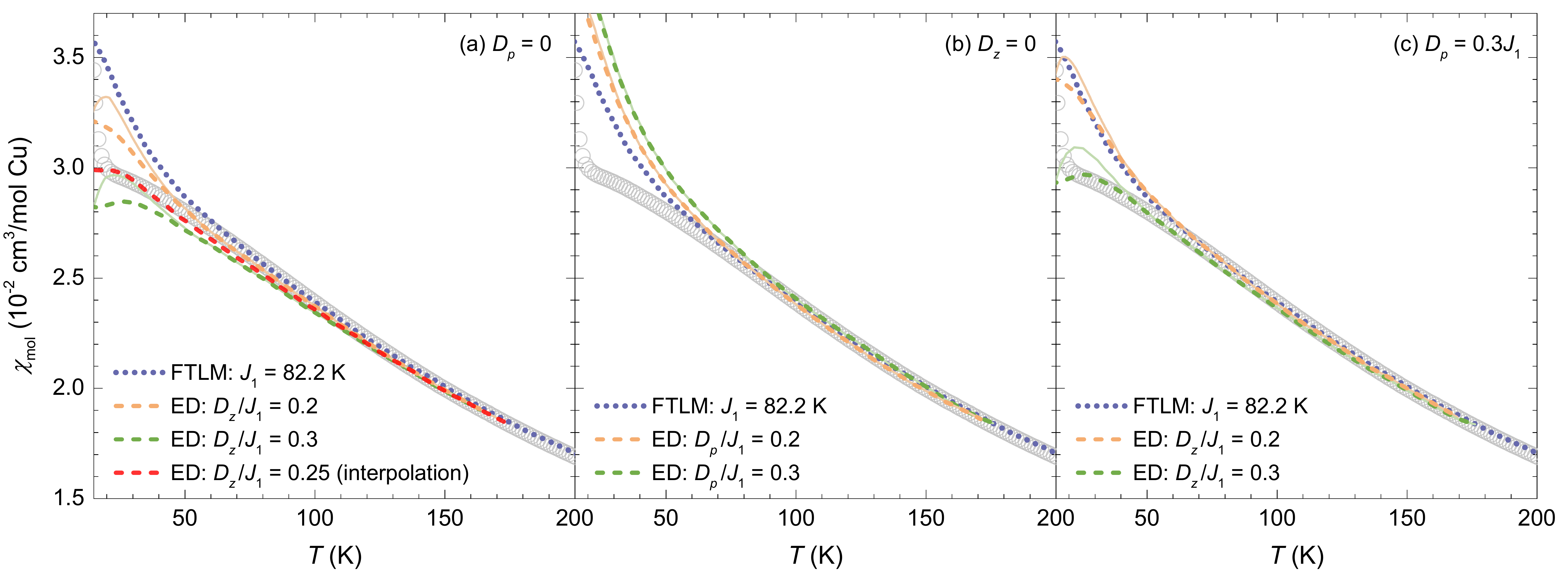}
\caption{ 
Molar susceptibility of {\YCu} in a field of 0.1\,T (circles) \cite{zorko2019YCu3muonSI}. 
The dotted line shows the FTLM prediction for the isotropic nearest-neighbor KAFM ($D_z = 0$) for $N=42$ sites \cite{schnack2018magnetismSI}. 
The solid and dashed lines correspond to ED calculations for an anisotropic model including additional Dzyaloshinskii-Moriya interaction for $N=12$ and 15 sites, respectively \cite{rigol2007kagomeSI}. 
The cases of finite out-of-plane DM component $D_z$ are shown in (a), finite in-plane component $D_p$ in (b) and both components being finite in (c).
The curve for $D_z/J_1=0.25$ and $D_p=0$ is obtained by interpolating between the published data.}
\label{figS2}
\end{figure*}

The molar susceptibility $\chi_{\rm mol}$ is calculated from susceptibility per site $\chi$ as
$
\chi_{\rm mol}=4\chi C/J_1,
$
where $C=\mu_0 (g\mu_B)^2 N_A /(4k_B)= 5.08$\,cm$^3$K/(mol~Cu) is the Curie constant, $k_B$ the Boltzmann constant, $\mu_B$ the Bohr magneton, $N_A$ the Avogadro number, and $g = 2.077$ the $g$-factor from the Curie-Weiss analysis shown in the inset in Fig.\,1 in the main text.
In Fig.\,\ref{figS2} various theoretical predictions for $\chi$ are compared to the experiment \cite{zorko2019YCu3muonSI}.
This includes FTLM calculations of isotropic nearest-neighbor KAFM on $N=42$-site spin clusters, which are cluster-size independent down to temperatures $T/J_1\simeq0.08 = 6.4\,{\rm K} \ll T_{\rm N}$ \cite{schnack2018magnetismSI}.
For the anisotropic model, which includes an additional Dzyaloshinskii-Moriya term ${\bf D} \cdot ({\bf S}_i\times {\bf S}_j)$ between nearest neighbors, we use the ED results of Rigol and Singh \cite{rigol2007kagomeSI} on $N=12$ and 15-site spin clusters.
For a sizable out-of-plane DM component $D_z$ these become size-dependent at $T\lesssim 0.4 J_1$ (Fig.\,\ref{figS2}a,c), while there is no size dependence down to $T \approx 0.2 J_1$ for a finite in-plane component $D_p$ (Fig.\,\ref{figS2}b).
However, even in the former case, the size-dependence remains below 4\% down to $T\lesssim0.2 J_1$, therefore, we use the more accurate $N=15$ results all the way down to $\sim$15\,K, where the experimental susceptibility starts increasing due to magnetic ordering.
As long as one of the DM components is zero, the sign of the other component does not affect the susceptibility curves.
However, once both are finite, the sign of $D_z$ makes a pronounced difference \cite{rigol2007kagomeSI}.
Here we only show results for $D_z>0$, which is the only choice compatible with the experimentally observed negative-chirality magnetically ordered ground state \cite{zorko2019negativeSI}.

\section{Electron Spin Resonance}

\begin{figure}[b]
\includegraphics[trim = 0mm 0mm 0mm 0mm, clip, width=1\linewidth]{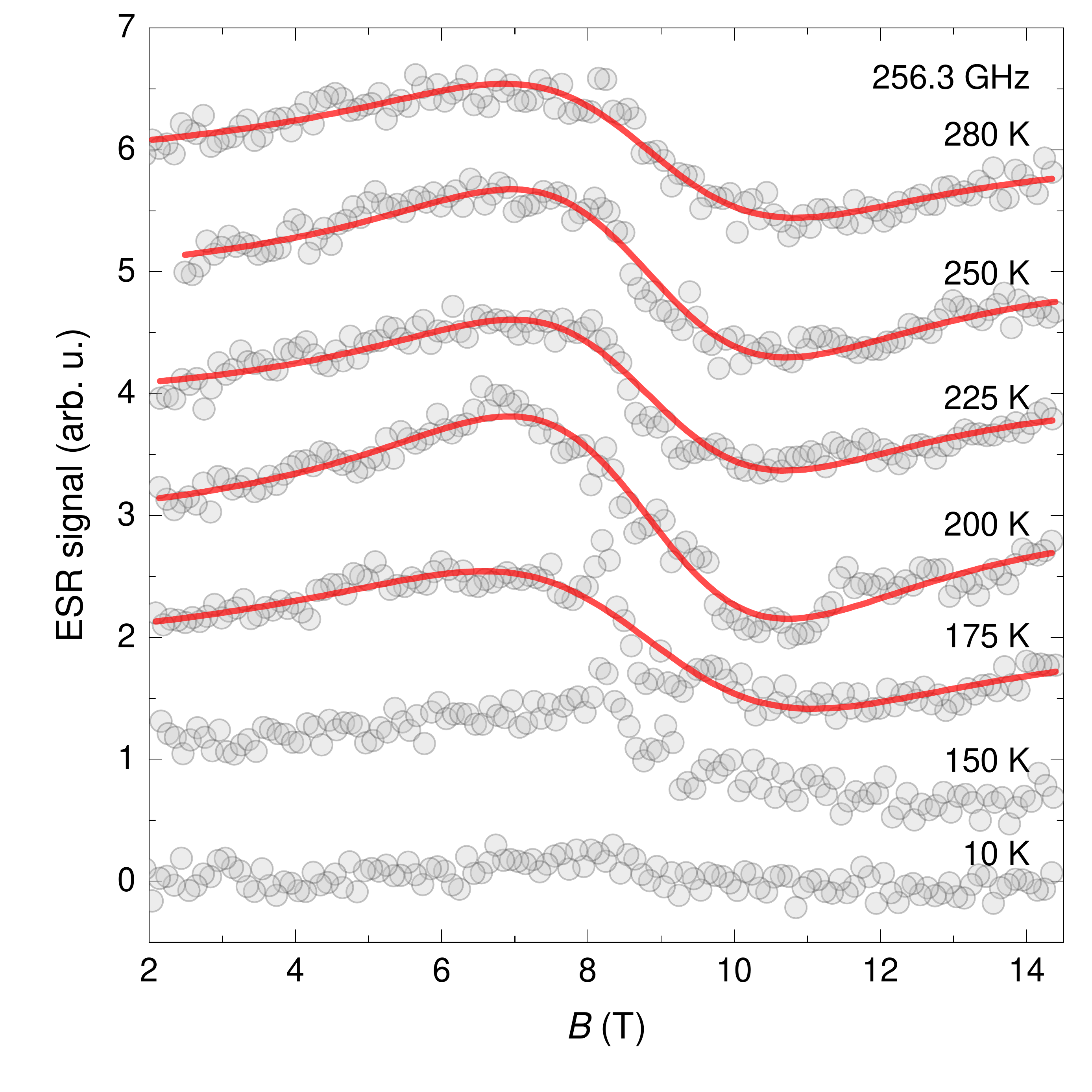}
\caption{The temperature dependence of the ESR spectra of {\YCu} measured at 256.3\,GHz (circles) with the corresponding Lorentzian fits (solid lines).
The spectra are offset vertically for clarity.}
\label{figS1}
\end{figure}
Electron spin resonance (ESR) measurements were performed at the National High Magnetic Field Laboratory, Tallahassee, USA on a custom-made transmission-type ESR spectrometer with homodyne detection equipped by a sweepable 15-T superconducting magnet.
The measurements were performed in the Faraday configuration at the irradiation frequencies of 212.6 and 256.3\,GHz on a 770-mg sample from the same batch as used in our previous investigations \cite{zorko2019YCu3muonSI, zorko2019negativeSI}.
A modulation field of about 2\,mT was used to record the derivative spectra.
The ESR spectra are a mixture of absorption and dispersion, with the corresponding phase determined from fits with the Lorentzian line shape.
In order to show pure absorption spectra, the spectra shown in Fig.\,\ref{figS1} are phase corrected.
The Lorentzian line shape of the ESR spectra is a sign of exchange narrowing due to strong exchange interactions \cite{zorko2018determinationSI}.
When fitting the spectra, we fixed the $g$ factor to $g=2.077$, as deduced from the Curie-Weiss analysis, because the spectra are too broad for a reliable $g$-factor determination.
The spectra are also too broad to resolve the angular dependence of the linewidth through powder spectra simulations, as was previously done for herbertsmithite \cite{zorko2008dzyaloshinskySI}.
Therefore, our fits can only yield a powder-averaged linewidth.
At $T\leqslant 150$\,K the fits become unreliable, since the spectra disappear  
in the noise.

For the powder-averaged full width at half maximum (FWHM) we 
use the well-established expression based on the calculation of the second and the fourth moment of the ESR line  \cite{zorko2008dzyaloshinskySI,zorko2013dzyaloshinskySI},
\begin{widetext}
\begin{equation}
\Delta B(\theta) = \sqrt{2 \pi} \frac{k_B}{2 g \mu_B J_1}
\sqrt{ \frac{\left[2D_z^2+3D_p^2+(2D_z^2-D_p^2){\rm cos}^2\theta\right]^3}{16D_z^2+78D_p^2+(16D_z^2-26D_p^2)\;{\rm cos}^2\theta}},
\label{eqDM}
\end{equation}
\end{widetext}
where $\theta$ denotes the angle between the magnetic field and the normal to the kagome planes and we consider only the dominant nearest-neighbor isotropic interaction.
By powder-averaging Eq.\,(\ref{eqDM}) and taking $g=2.077$, $J_1=82$\,K and the experimental high-temperature linewidth $\Delta B = 6.8$\,T we obtain the solution $D_z(D_p)$ shown in Fig.\,3 in the main text.

\begin{figure}[b]
\includegraphics[trim = 0mm 0mm 0mm 0mm, clip, width=1\linewidth]{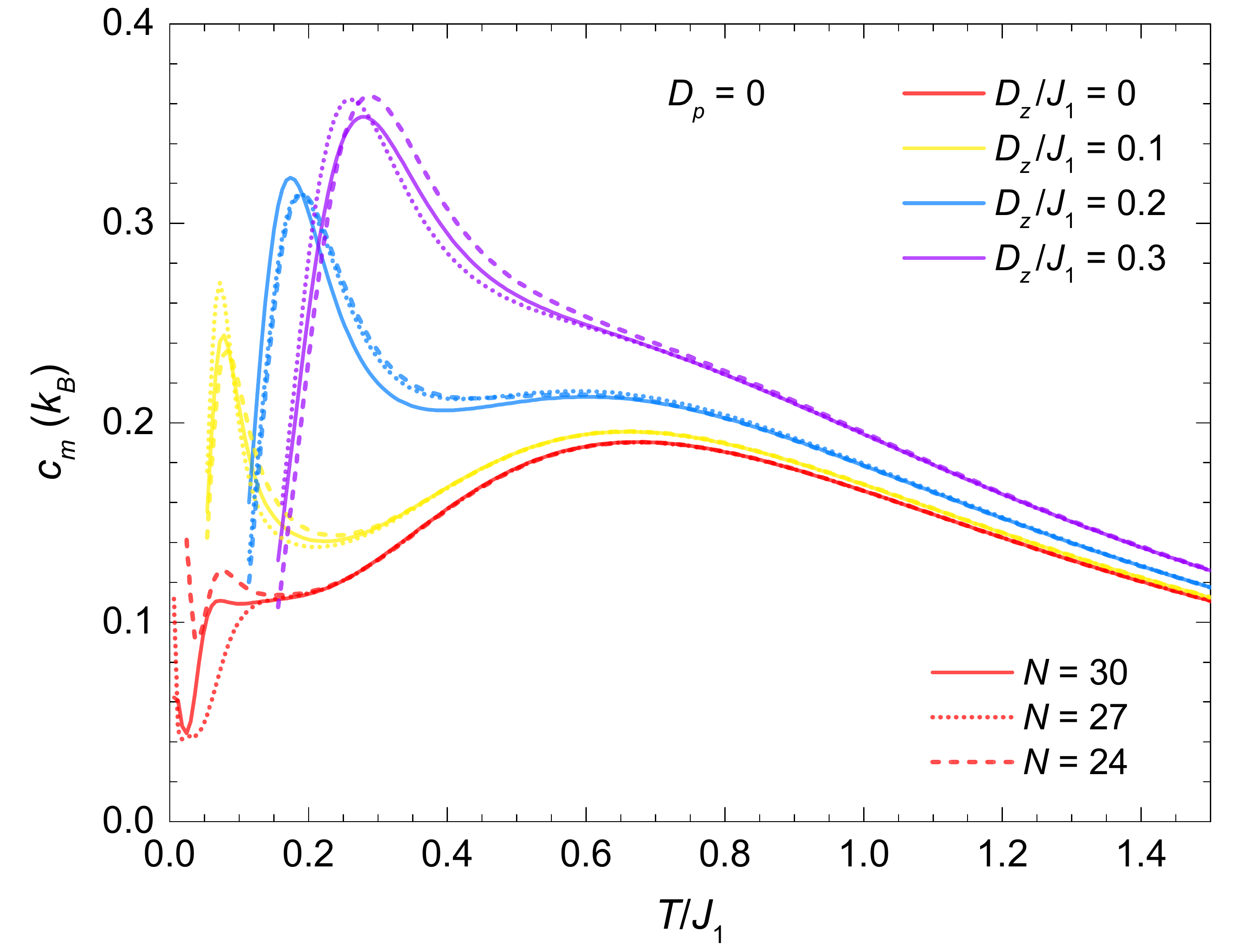}
\caption{The temperature dependence of the magnetic specific heat per spin in zero magnetic field obtained from FTLM calculations  for various sizes ($N$) of spin clusters and DM coupling values $D_z/J_1$ ($D_p=0$). The data are shown for temperatures where $s\geq 0.07 k_B$.
}
\label{figS3}
\end{figure}

\section{FTLM Calculations}

We consider the spin-1/2 model on the kagome lattice with the Heisenberg nearest-neighbor exchange and additional Dzyaloshinskii-Moriya (DM) interactions.
The corresponding Hamiltonian is
 \begin{equation} 
 \mathcal{H}=  \sum_{\langle ij \rangle}  \left[ J_1 {\bf S}_i \cdot {\bf S}_j+ {\bf D}_{ij}
 \cdot  ({\bf S}_i \times {\bf S}_j) \right].
 \label{model}
 \end{equation} 
The DM anisotropy term within the kagome lattice involves (in general) two independent parameters, the component perpendicular to the kagome plane $D^z_{ij} =  D_z$ and the in-plane component $|D^p_{ij}| = D_p$, which is perpendicular to the bond and is also allowed in {\YCu} because the kagome plane is not a mirror plane.
We use the convention of counting the bonds in all triangles in the same (counterclockwise) rotation yielding the DM vector pattern shown in Ref.\,\cite{zorko2008dzyaloshinskySI}.
As shown later on, specific heat (and entropy) is mostly sensitive on $D_z$, while the effect of $D_p$ is almost negligible.

Within this model we calculate the entropy per site $s(T)$, and consequently the specific heat $c_m(T)= T {\rm d}S/{\rm d}T$, using the finite-temperature Lanczos method (FTLM) \cite{jaklic94finiteSI,jaklic00finiteSI}, previously used in numerous studies of static (and dynamical) properties at $T>0$ in various models of correlated electrons \cite{prelovsek13stronglySI}, including thermodynamic quantities of the pure Heisenberg model
on the kagome lattice \cite{schnack2018magnetismSI}. 
As the $D_p=0$ but $D_z \neq 0$ model still retains the conservation of $S^z_{\rm tot}$ and the translational symmetry (due to periodic boundary conditions),  the memory and CPU time requirement for a given system size $N$ are essentially that of the Lanczos procedure for the ground state, provided that we scan over all (different) symmetry sectors $S_{\rm tot}^z$ and wavevectors $q$, and in addition perform a modest sampling over the initial wavefunctions with $N_s \sim 30$. 
In the present study we thus deal with the kagome lattices with up to $N=30$ sites, where the biggest symmetry sector contains $N_{\rm st} \sim 16 \times 10^6$ basis states.

\begin{figure}[b]
\includegraphics[trim = 0mm 0mm 0mm 0mm, clip, width=1\linewidth]{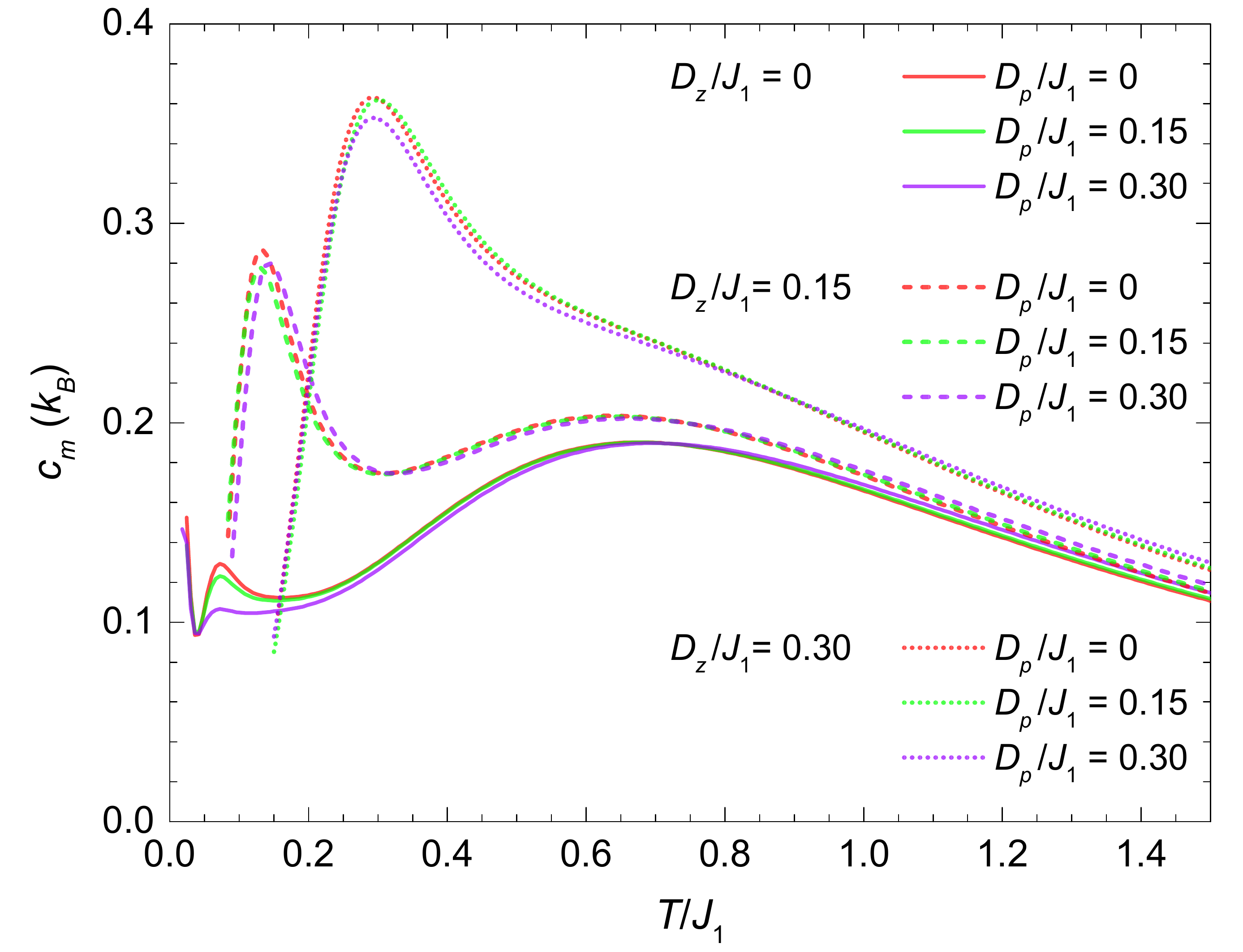}
\caption{The temperature dependence of the magnetic specific heat per spin in zero magnetic field obtained from FTLM calculations on $N=24$-site spin clusters for various $D_z/J_1$ and $D_p/J_1$ values. The data are shown for temperatures where $s\geq 0.07 k_B$.}
\label{figS5}
\end{figure}

While FTLM is quite accurate for a given finite-size system, the main concern is the macroscopic ($N \to \infty$) validity of the obtained results \cite{jaklic94finiteSI,jaklic00finiteSI}.
Typically, the criterion $T>T_{\rm fs}$ is related to the grandcanonical sum
$Z(T) = \mathrm{Tr}
\{\exp[-(\mathcal{H}-E_0)/T\}$, where $E_0$ is the ground-state energy and we require $Z > Z(T_{\rm fs}) \gg 1$.
Since $Z(T)$ is closely related to entropy $s(T)$, this requirement in actual systems effectively reduces to $s > (0.07 - 0.1) k_B$. Fortunately, frustrated systems are characterized by large $s(T)$ at low temperatures and consequently $T_{\rm fs} \ll J_1$.
This is particularly the case for the pure Heisenberg ($D_z =0$) model on the kagome lattice (Fig.\,4b in the main text; see also the finite-size analysis in Ref.\,\cite{schnack2018magnetismSI}).
To demonstrate that finite-size effects are small in the considered temperature range where $s(T)\geq 0.07$, we present in Fig.\,\ref{figS3} the comparison of FTLM results for $c(T)$ obtained on lattices with $N =24, 27$, and 30 sites.
The finite-size effects are somewhat enhanced only for the $D_z=0$ case at $T/J_1<0.15$.

We can treat also the case of $D_p \ne 0$ by FTLM, but this
requires to abandon $S^z_{\rm tot}$ as the conserved quantity. 
Consequently we can only work with smaller finite-size systems (up to $N=24$).
In Fig.\,\ref{figS5}, $c_m$ is shown for various combinations of $D_z$ and D$_p$. 
The influence of $D_p$ is very small for modest values of $D_p/J_1 \leq 0.3$ and $T/J_1 \geq0.15$, in sharp contrast to the strong effect of $D_z$. 
In particular, $D_p$ does not affect the position of the low-temperature maximum in $c_m$, which is thus a unique fingerprint of the out-of-plane DM interaction $D_z$. 

\section{Modeling of specific heat}

\begin{figure}[t]
\includegraphics[trim = 0mm 0mm 0mm 0mm, clip, width=1\linewidth]{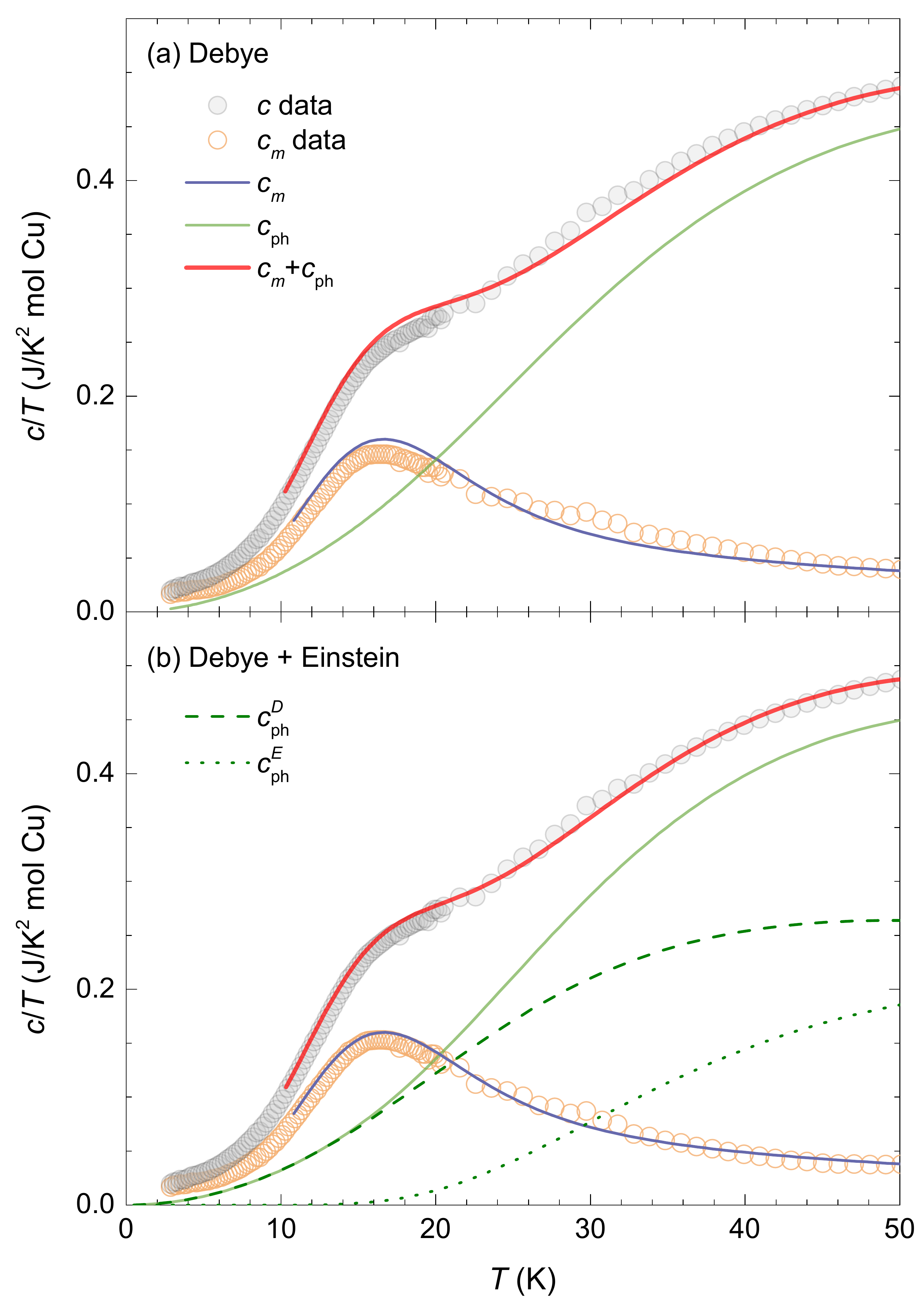}
\caption{The fit of the specific heat data (symbols) with the model $c=c_m+c_{\rm ph}$. 
The theoretical prediction for the magnetic part $c_m$ corresponds to FTLM calculations for $D_z/J_1 = 0.25$ and the phonon part to (a) a pure Debye contribution or (b) a sum of a Debye and an Einstein contribution.
The $c_m$ data is obtained by subtracting the theoretical phonon contribution from the measured specific heat.}
\label{figS4}
\end{figure}

The total specific heat of the sample is a sum of
the magnetic contribution $c_m$ and a phonon contribution $c_{\rm ph}$, $c=c_m+c_{\rm ph}$. 
For the former, we use FTLM calculations from the previous section for $D_z/J_1=0.25$. 
This value of the anisotropy is determined by the position $T_{\rm max}$ of the experimentally observed low-temperature maximum in $c/T$ (Fig.\,4c in the main text).
For the phonon contribution we first use a pure Debye model  \cite{tari2003specificSI}
\begin{equation}
c_{\rm ph}^D \propto \left(\frac{T}{\theta_D} \right)^3 \int_0^{\,\theta_D/T}{\frac{x^4 {\rm e}^x {\rm d}x}{\left({\rm e}^x - 1 \right)^2}}.
\end{equation} 
The fit to the experimental data is quite good (Fig.\,\ref{figS4}a) and yields the Debye temperature $\theta_D=224(5)$\,K.
An even better fit is obtained if an additional optical mode is taken into account (Fig.\,\ref{figS4}b), yielding the Einstein contribution to the specific heat \cite{tari2003specificSI}
\begin{equation}
c_{\rm ph}^E \propto \left(\frac{\theta_E}{T} \right)^2\frac{{\rm e}^{\theta_E/T}}{\left({\rm e}^{\theta_E/T} - 1 \right)^2}.
\end{equation} 
Since the position of optical phonon modes is not known for \YCu, we fix the Einstein temperature to $\theta_E=177$\,K, which corresponds to the lowest
Raman active mode found at 123\,cm$^{-1}$ in the structurally similar herbertsmithite \cite{wulferding2010interplaySI}.
In this second model the Debye temperature is decreased to $\theta_D=177(5)$\,K.

We stress that the selection of the specific phonon model does not affect the ultimate determination of the dominant DM component $D_z$, although the model with the additional Einstein mode does give a somewhat better agreement between the theoretically predicted and the experimentally observed $c_m$ (Fig.\,\ref{figS4}).
This insensitivity to the phonon model is due to the fact that both the Debye and the Einstein contributions to  $c/T$ are characterized by maxima at $T\gg T_{\rm max}$ (see Fig.\,\ref{figS4}).
Consequently, the experimentally determined magnetic contribution to the specific heat is well-defined irrespective of the phonon model.
The low temperature maximum at $T_{\rm max} = 16$\,K is, therefore, a stringent measure of the DM interaction on the kagome lattice.  
The high-temperature maximum in specific heat around $T=0.67J_1= 55$\,K (see Fig.\,4a in the main text) is, on the other hand, much harder to confirm experimentally. 
It is heavily affected by the phonon model, because at 55\,K the phonon contribution overshadows the magnetic contribution as $c_{\rm ph}/c_m \approx 15$.
Nevertheless, the presence of this maximum is not in question, as it is not lattice dependent, i.e, it appears at $T = 0.67 J$ also on a square lattice  \cite{sengupta2003specificSI}.
The reason for its universality is that it corresponds to the development of the nearest-neighbor spin correlations, which appear at temperatures substantially higher than those at which the details of the specific spin lattice would start to become important. 
The maximum appears at temperatures where the spin correlation length is on the order of the lattice spacing, and is thus not yet large enough for spins to show a coherent response beyond nearest-neighbor pairs.

\section{Short-range vs. long-range ordering}

\begin{figure}[t]
\includegraphics[trim = 0mm 0mm 0mm 0mm, clip, width=1\linewidth]{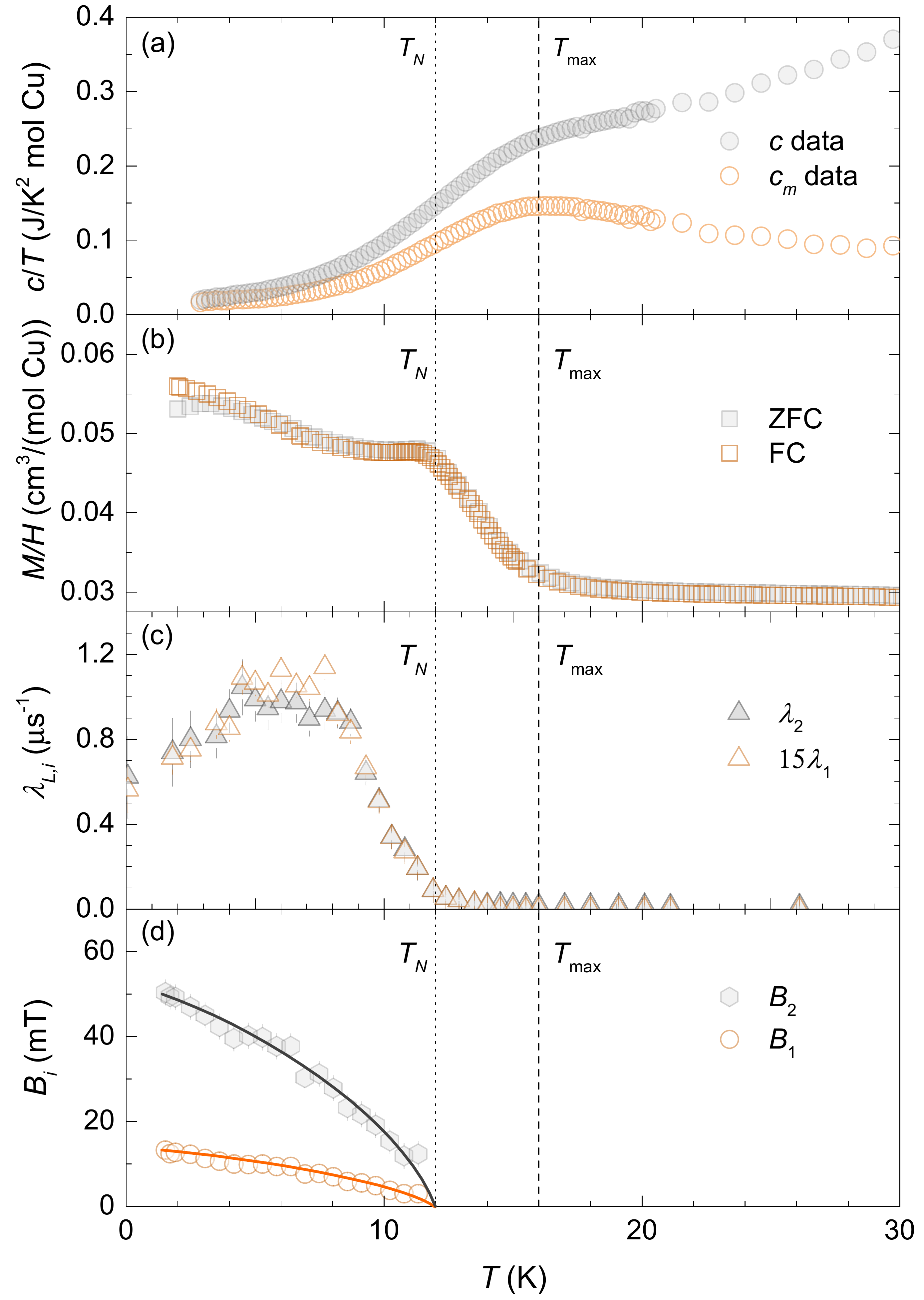}
\caption{The difference between the on-set temperature $T_{\rm max}$ of 2D short-range chiral spin correlations, determined from (a) the low-temperature maximum in magnetic specific heat, and the on-set temperature $T_N$ of 3D long-range order, determined from (b) a cusp in magnetic susceptibility, (c) the enhancement of longitudinal muon spin relaxation and (d) the appearance of static internal fields. The data in panels (b-d) are taken from Ref.\,\cite{zorko2019YCu3muonSI}.
}
\label{figS6}
\end{figure}

As deduced from the FTLM calculations, the low-temperature maximum in specific heat at $T_{\rm max}=16$\,K corresponds to the establishment of short-range chiral spin correlations within the kagome layers.
3D long-range ordering is established at a lower temperature of $T_N=12$\,K, as detected by several experimental observables shown in Fig.\,\ref{figS6}.
This includes a cusp in bulk magnetic susceptibility (Fig.\,\ref{figS6}b), sudden enhancement of longitudinal muon spin relaxation (Fig.\,\ref{figS6}c), and the appearance of static internal fields (Fig.\,\ref{figS6}d) \cite{zorko2019YCu3muonSI}. 

\end{document}